\documentclass[twocolumn,showpacs,amsmath,amssymb,bibnotes]{revtex4}
\usepackage{bm}
\usepackage{graphicx,psfig,epsfig,subfigure,afterpage}
\usepackage{amsfonts}

\newcommand{\be}{\begin{equation}}
\newcommand{\ee}{\end{equation}}
\newcommand{\bea}{\begin{eqnarray}}
\newcommand{\eea}{\end{eqnarray}}
\newcommand{\gdot}{\dot{\gamma}}
\newcommand{\gdotbar}{\overline{\dot{\gamma}}}

\newcommand{\etal}{{\it et al.\/}}
\newcommand{\bw}{\begin{widetext}}
\newcommand{\ew}{\end{widetext}}

\newcommand{\vecv}[1]{\mathbf{{#1}}}
\newcommand{\tens}[1]{\mathbf{{#1}}}
\newcommand{\nablu}{{\bf \nabla}}

\begin{document}

\title{Nonlinear dynamics of a shear banding interface}
\author{S. M. Fielding}
\email{suzanne.fielding@manchester.ac.uk}
\affiliation{School of Mathematics, University of Manchester, Booth
  Street East, Manchester M13 9EP, United Kingdom } 
 \author{P. D. Olmsted}
\affiliation{Polymer IRC and School of Physics \& Astronomy,
  University of Leeds, Leeds LS2 9JT, United Kingdom}  
\date{\today}
\begin{abstract}
  
  We study numerically the nonlinear dynamics of a shear banding
  interface in two dimensional planar shear flow, within the non-local
  Johnson Segalman model. Consistent with a recent linear stability
  analysis, we find that an initially flat interface is unstable with
  respect to small undulations for sufficiently small ratio of the
  interfacial width $\ell$ to cell length $L_x$. The instability
  saturates in finite amplitude interfacial fluctuations. For
  decreasing $\ell/L_x$ these undergo a non equilibrium transition
  from simple travelling interfacial waves with constant average wall
  stress, to periodically rippling waves with a periodic stress
  response. When multiple shear bands are present we find erratic
  interfacial dynamics and a stress response suggesting low
  dimensional chaos.
\end{abstract}
\pacs{{47.50.+d}, 
     {47.20.-k}, 
     {36.20.-r}.
     } 
\maketitle


Complex fluids such as polymers, liquid crystals and surfactant
solutions have mesoscopic structure that is readily perturbed by flow
\cite{LarsonComplex}.  For example, wormlike surfactant micelles with
lengths of the order of microns can be induced to stretch, disentangle and
entangle, and break or increase in length.  Their mechanical response
is therefore highly non-Newtonian, with shear flows inducing normal
stresses (\textit{e.g.}  $\sigma_{xx}-\sigma_{yy}$), and with the
shear stress $\sigma_{xy}$ being a nonlinear function of applied shear
rate $\gdot$.  Recent work on such fluids has led to a fairly
consistent picture of ``shear banding'': coexistence in shear flow of
viscously thicker (nascent) and thinner (flow-induced) bands of
material flowing at different local shear rates, for an overall
average imposed shear rate.

This phenomenon can be described by constitutive models for which the
shear stress is a non-monotonic function of shear rate for homogeneous
flow. This leads to a separation into bands of differing shear rate
that coexist at a common shear stress \cite{SCM93b,olmsted99d}.
Although most studies have assumed a flat interface between the bands
and (with a few exceptions \cite{fielding04}) predicted a
time-independent banded state, an accumulating body of data
has demonstrated that the average shear stress, and the banding
interface, can fluctuate
\cite{Becu.Manneville.ea04,Lopez-Gonzalez.Holmes.ea04,WunColLenArnRou01,Holmes.Lopez-Gonzalez.ea03,HBP98,WFF98,Manneville.Salmon.ea04,chaos2000}.
An important question is then whether these fluctuations resemble
small amplitude capillary waves stabilised by surface tension, or
whether they arise from an underlying instability, stabilised at large
amplitude by nonlinearities.  In this Letter we give strong evidence
supporting the latter scenario, via the first theoretical study of the
nonlinear dynamics of a shear banding interface.

\textit{The model -- } The generalised Navier Stokes equation
for a viscoelastic material in a Newtonian solvent of viscosity $\eta$
and density $\rho$ is:
\begin{equation}
\label{eqn:NS}
\rho(\partial_t + \vecv{v}.\nablu)\vecv{v} = \nablu .(\tens{\Sigma} +
2\eta\vecv{D} -P\tens{I}), 
\end{equation}
where $\vecv{v}(\vecv{r})$ is the velocity field. The pressure $P$ is
determined by incompressibility, $\vecv{\nabla}\cdot\vecv{v}=0$.  The
viscoelastic stress $\vecv{\Sigma}(\vecv{r})$ evolves according to the
non-local (``diffusive'') Johnson Segalman (DJS)
model~\cite{johnson77,olmsted99a}
\begin{gather}\label{eqn:DJS}
(\partial_t
+\vecv{v}\cdot\nablu )\,\tens{\Sigma} 
- a(\tens{D}\cdot\tens{\Sigma}+\tens{\Sigma}\cdot\tens{D}) \\- 
(\tens{\Sigma}\cdot\tens{\Omega} - \tens{\Omega}\cdot\tens{\Sigma})  
 = 2 G\tens{D}-\frac{\tens{\Sigma}}{\tau}+ \frac{\ell^2}{\tau }\nablu^2 
 \tens{\Sigma}, \nonumber
\end{gather}
with plateau modulus $G$ and relaxation time $\tau$. $\tens{D}$ and
$\tens{\Omega}$ are the symmetric and antisymmetric parts of the
velocity gradient tensor, $(\nablu \vecv{v})_{\alpha\beta}\equiv
\partial_\alpha v_\beta$.  For $a=1$ and $\ell=0$ this model reduces
to the Oldroyd B model, which is motivated by considering an ensemble
of beads paired by springs (simplified polymer chains) into dumbbells.
Stress is generated as the flow deforms the dumbbells, and is relaxed
on the timescale $\tau$ for the springs to regain equilibrium length.
To capture shear thinning the DJS model invokes a ``slip parameter''
$a$ with $|a|<1$ to give non-affine dumbbell
deformation~\cite{johnson77}. The constitutive curve
$T_{xy}=\Sigma_{xy}(\gdot, a)+\eta\gdot$ for homogeneous planar shear
$\vecv{v}=y\gdot\vecv{\hat{x}}$ is then capable of non-monotonicity,
allowing shear banding. The non-local diffusive term in
Eqn.~\ref{eqn:DJS} accounts for spatial gradients across the banding
interface on a length scale $\ell$ \cite{olmsted99a}. It arises
naturally in models of liquid crystals, and diffusion of polymer
molecules~\cite{elkareh89}. In the context of one dimensional (1D)
calculations, it has been shown to give a unique (selected) stress
$T_{xy}^*$ at which banding occurs~\cite{olmsted99a}, as seen
experimentally.

We study 2D boundary-driven flow between parallel plates at $y=0,L$
with plate conditions
$\partial_y\Sigma_{\alpha\beta}=0\;\forall\;\alpha,\beta$ for the
viscoelastic stress, and no slip or permeation for the velocity.  In
the flow direction we consider a domain $x=\{0,L_x\}$, with periodic
boundary conditions.  We choose $a=0.3, \eta=0.05$ throughout, and
units in which $G=1,\tau=1$ and $L=1$.

For an imposed average shear rate $\gdotbar\equiv[v_y(L)-v_y(0)]/L$ in
the region of decreasing stress, $d T_{xy}/d\gdot<0$, homogeneous flow
is unstable~\cite{Yerushalmi70}. A 1D ($y$) calculation then predicts
separation into bands of shear rates $\gdot_1=0.66, \gdot_2=7.09$, at
a selected shear stress $T^{\ast}_{xy}=0.506$.  Recent
analysis~\cite{SMF2005} showed this stationary 1D banded state to be
linearly unstable to 2D ($x, y$) perturbations corresponding to
undulations of the interface with wavevector
$\vecv{q}=q_x\vecv{\hat{x}}$. The most unstable mode has $q_xL\approx
2\pi$, and the instability involves feedback of the normal stress with
velocity fluctuations across the interface. In this Letter, we study
the fate of the interface in the \textit{nonlinear} regime, and
demonstrate it to be restabilised at the level of finite amplitude
undulations.

\begin{figure}[htb]
  \centering
\includegraphics[scale=0.3]{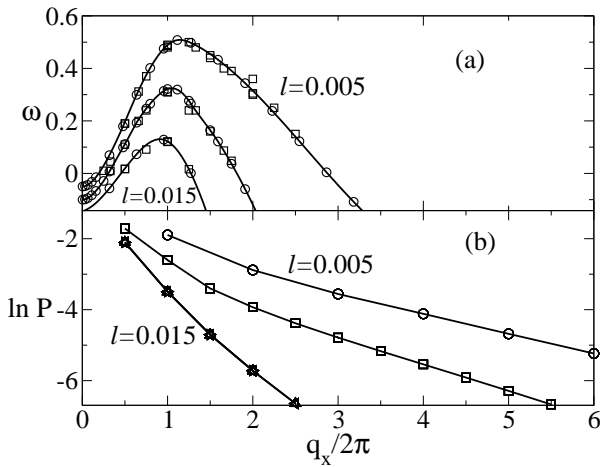}
 \caption{(a) Growth rate $\omega(q_x)$ of perturbations about an
   initially 1D banded state. ($\circ$, \textbf{---}): predictions of
   linear stability \cite{SMF2005} for $\ell=0.005,\, 0.01,\, 0.015$.
   ($\Box$): early time dynamics of the full 2D numerics for
   $\ell=0.005,\,L_x=4,\,6,\,8$; $\ell=0.01,\,L_x=2,\,4,\,6,\,8$;
   $\ell=0.015,\,L_x=2,4,8$. (b) Power spectrum $P(q_x)$ in the
   travelling wave regime for $\ell=0.005,\,L_x=1$ ($\circ$);
   $\ell=0.01,\,L_x=2$ ($\Box$); $\ell=0.015,\,L_x=2,\,4,\,6$
   ($\triangle$).}
 \label{fig:dispersion}
\end{figure}
\begin{figure}[t]
  \centering
\includegraphics[scale=0.9]{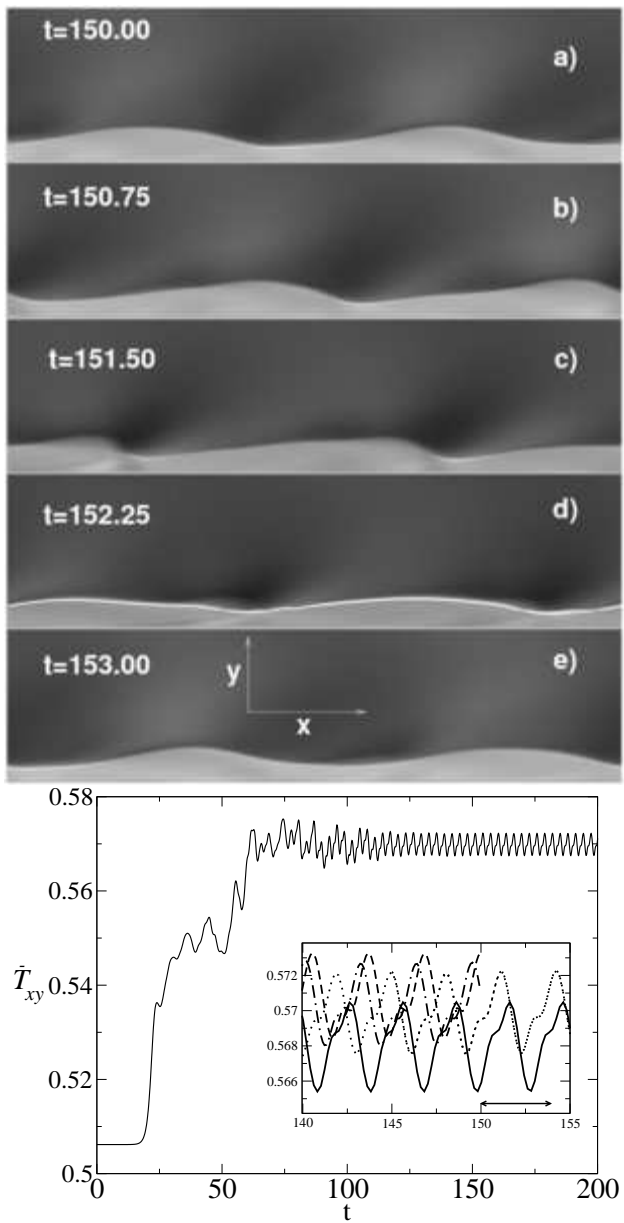}
\includegraphics[scale=0.3]{./stress.eps} 
\caption{Rippling wave  at $\ell=0.005, L_x=4, \gdotbar=2$.  Top:
  Greyscale of $\Sigma_{xx}(x,y)$.  Upper wall moves to the
  right. White line in d): interface height defined in text. Bottom:
  average wall stress. Inset: $(N_x,N_y,dt)=(200,800,0.0003)$, solid
  line; $(400,800,0.0001)$, dotted; $(400,800,0.0003)$, dashed;
  $(500,800,0.0002)$, dot-dashed.}
\label{fig:break}
\end{figure}

\textit{Numerical scheme---} We enforce incompressibility by
eliminating the velocity in favour of a stream-function $\psi$. At
each time step we update the local parts of Eqn.~\ref{eqn:DJS} using
an explicit Euler algorithm within a finite difference scheme on a
rectangular grid of $(N_x, N_y)$ nodes, using third order upwinding
for the convective term $(\vecv{v}.\nablu)\tens{\Sigma}$. We then take
a Fourier transform $x\to q_x$ in the flow direction and update the
nonlocal part of Eqn.~\ref{eqn:DJS} using the semi-implicit
Crank-Nicolson algorithm. We finally update Eqn.~\ref{eqn:NS} at zero
Reynolds number, $\rho=0$. We attain convergence to one percent with
respect to increasing spatial and temporal resolution. This ensures
that the results shown are converged to the eye, apart from the stress
signal $\bar{T}_{xy}(t)$ in Fig.~\ref{fig:break}: this shows slight
quantitative, but not qualitative, changes. Runs at finer resolution
are prohibitively time consuming. We study two initial conditions:
(IC1) the stationary banded state, predicted by a 1D calculation, of
two bands separated by flat interface; and (IC2) a homogeneous
unstressed fluid, corresponding to shear startup from rest.

\begin{figure}[htb]
  \centering
\includegraphics[scale=0.3]{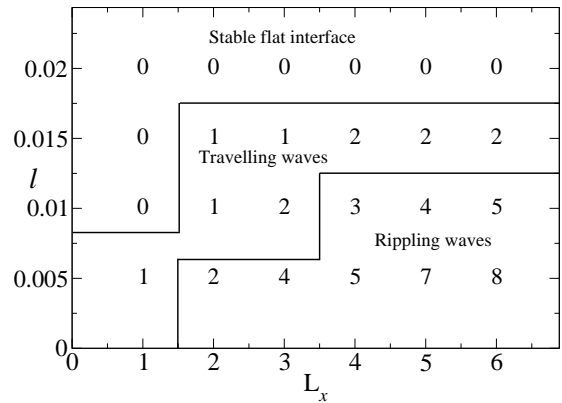}
 \caption{``Phase diagram''  for 
   $\bar{\gdot}=2.0$, showing the different non-linear regimes for IC1
   (accurate to  spacing between numbers) and the numbers of
   linearly unstable modes (Ref.~\cite{SMF2005}). }
 \label{fig:phasediag}
\end{figure}
\textit{Linear regime---} First we consider the early time evolution
of an initially flat interface, IC1. For a wide enough interface
$\ell$ and small enough system size $L_x$ this 1D solution is stable.
For smaller $\ell/L_x$, however, interfacial undulations are predicted
by the linear analysis of Refs.~\cite{SMF2005} to become unstable. Our
numerics successfully reproduce this instability during the initial
evolution away from IC1: the eigenvectors and growth rates
$\omega(q_x)$ match the analytical results of Refs.~\cite{SMF2005},
Fig~\ref{fig:dispersion}.
 Beyond this early-time regime,
the interface is restabilised by nonlinear effects, and the
instability saturates in undulations of finite large amplitude.
Depending on the distance from the onset of linear instability,
Fig.~\ref{fig:phasediag}, the ultimate attractor corresponds either to
a steady travelling interfacial wave, or to periodically rippling
waves.

\begin{figure}[b]
  \centering
\includegraphics[scale=0.95]{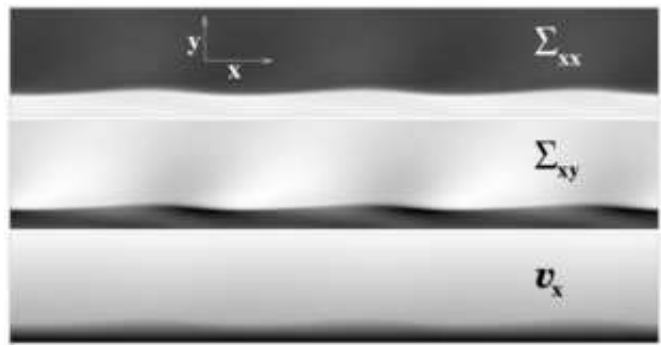}
 \caption{Greyscale of order parameters for travelling wave
   in the $(x,y)$ plane for $\ell=0.015$, $L_x=6$, and upper wall
   velocity $V\equiv \gdotbar L=2$ to the right.
\label{fig:travel}}
\end{figure}

\textit{Travelling wave---} For values of $\ell/L_x$ marginally inside
the unstable regime (Fig.~\ref{fig:phasediag}) the ultimate attractor
comprises a travelling wave $\vecv{A}=\vecv{A}(y,x-ct)$ for all order
parameters $\vecv{A}=\{\Sigma_{xx},\Sigma_{xy},\Sigma_{yy},\psi\}$.
Since this satisfies periodic boundary conditions, the wall-averaged
shear stress is constant in time, with a value
$\overline{T}_{xy,ss}\simeq0.51-0.54$ that depends on $\ell$ and $L_x$ and
is slightly higher than the selected stress $T_{xy}^{\ast}=0.506$ of
the 1D calculation. In Fig.~\ref{fig:travel} the wave-speed $c=1.15$,
compared with a horizontal velocity $v_x$ that varies between $1.439$
and $1.484$ along the interface.
To analyse the structure of this
state we define the following power spectrum,
\begin{equation}
  \label{eq:1}
   P(q_x)=\sum_{i=1}^4\int_0^L\,dy| A_i(q_x,y)|^2.
\end{equation}
%
%
%
\begin{figure}[t]
  \centering
\includegraphics[scale=0.45]{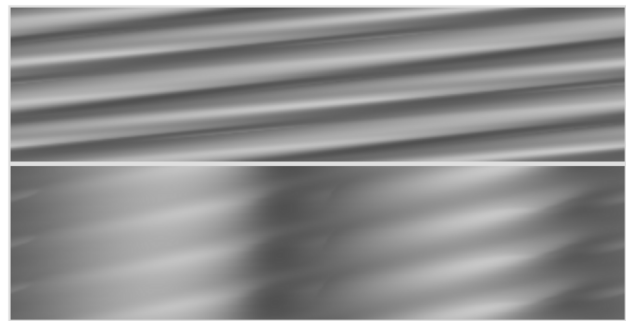}
 \caption{Rippling wave regime. Greyscale of interface height (defined
   in text) in  $(x,t)$ plane. Parameters as in
   Fig.~\ref{fig:break}. Time $t=150\ldots160$ upwards. Upper: raw
   data. Lower:  transformed as $x\to x-ct$, $c$ extracted by
   eye.}\label{fig:interface} 
\end{figure}
%
As seen in Figs.~\ref{fig:dispersion} and~\ref{fig:phasediag}, while
the linear stability analysis predicted only a few unstable modes, the
ultimate nonlinear state has active modes at higher $q_x$,
Fig.~\ref{fig:dispersion}b.  Interestingly, the dominant mode is still
close to that of the linear analysis, despite being in a nonlinear
regime with a finite interfacial displacement $\delta h$.
Indeed, for all cases in Fig. \ref{fig:dispersion}b, we find
empirically that the dominant mode has the longest wavelength that is
both consistent with periodic boundary conditions and linearly
unstable.


\textit{Rippling wave---} For smaller $\ell/L_x$, deeper inside the
unstable regime (Fig.~\ref{fig:phasediag}), we see a new regime in
which the travelling wave now periodically ``ripples''.  The
wall-averaged stress $\overline{T}_{xy}$ is periodic in time
(Fig.~\ref{fig:break}, again for IC1), with variations of the order of
one percent, and an average value larger than the 1D selected stress
$T_{xy}^{\ast}$. The interface height $h(x)$ is shown as a white line
in Fig.~\ref{fig:break}d and as a greyscale over a time window in
Fig.~\ref{fig:interface}.  Because the structure is quite complicated
near a rippling wave (Figs.~\ref{fig:break}cd), the location of $h(x)$
depends on its precise definition. We take $h(x)$ as that value of $y$
at which $\Sigma_{xx}$ lies half way between its values at the two
walls. (An alternative might be $\int dy\,y \,| \partial_y
\Sigma_{xx}(y)|$.)  This provides a fairly reliable measure, subject
to a small kink in the rippling region (Fig.~\ref{fig:break}d).

\begin{figure}[t]
  \centering
\includegraphics[scale=0.57]{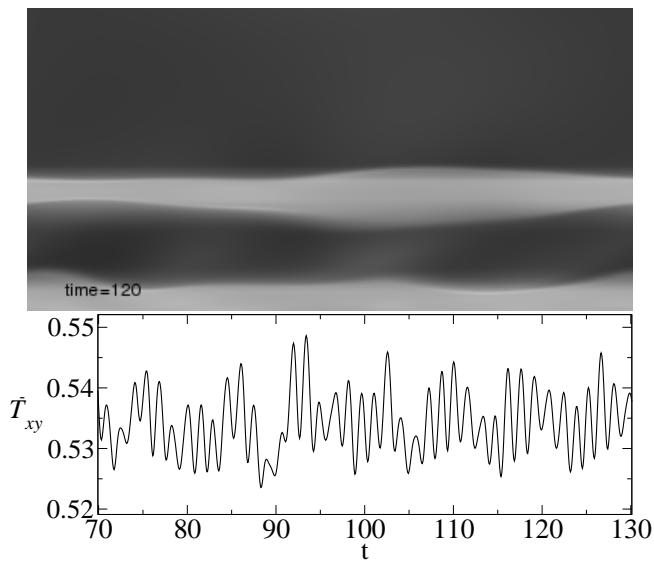}
\includegraphics[scale=0.29]{./multiple2.eps}
 \caption{Multiple bands after startup from rest for
   $\ell=0.005$, $L_x=2$, $\gdotbar=2$. Top: greyscale of
   $\Sigma_{xx}(x,y)$ at time $t=120$, reconstructed from the first
   $15$ Fourier modes. Bottom: Shear stress $T_{xy}(t)$.
   \label{fig:multiple1}}
\end{figure}

\textit{Robustness to initial conditions---} So far, we have
considered only IC1, two bands separated by a single flat interface.
To test robustness to ICs, we now study startup from rest (IC2).  Here
the system develops either (i) two bands that show the same dynamics
as with IC1, or (ii) multiple bands that have erratic dynamics
suggestive of low dimensional chaos, Fig.~\ref{fig:multiple1}.  This
is the counterpart of startup from rest in 1D planar shear flow, which
typically yields random configurations of multiple bands, because the
uniform shear stress $T_{xy}(y)=T^*_{xy}$ allows interfaces to reside
at any $y$ value. (To achieve IC1, the system was prebiased to form
two bands.)  In contrast, Couette flow between concentric cylinders
has a stress $T_{xy}\sim1/r^2$ ($r$=radius), which allows only a
single stationary interface at $r^*\sim1/\sqrt{T^*_{xy}}$, as
demonstrated in 1D in Ref.~\cite{GrecBall97,radulescu99a}. By analogy,
we anticipate that curvature should eliminate the multiple band case
(ii) in 2D.  Although a true Couette calculation has not been
performed, we implement a ``poor man's'' version by adding a biasing
stress-gradient in the $y$ direction while the bands are forming. For
strong enough bias, we then indeed find case (i), just two bands. An
open question is whether, even with IC1, a transition exists in the
numerically inaccessible regime of small $\ell/L_x$ (far bottom right
in Fig.~\ref{fig:phasediag}) to chaotic dynamics of a single
interface.

To summarise, we have studied numerically the nonlinear dynamics of a
shear banding interface in 2D planar shear flow, within the non-local
DJS model, for a system of length $L_x$ with periodic boundary
conditions in $x$. Upon decreasing the ratio of the interfacial width
$\ell$ to the system length $L_x$ the stable two band state separated
by a single interface undergoes successive transitions to travelling
waves and rippling travelling waves.  Multiple shear bands can also be
found, depending on initial conditions, which show irregular
interfacial dynamics and a corresponding stress signal suggestive of
low dimensional chaos.  Depending on the regime, this dynamics is
qualitatively similar to several recent experiments in wormlike
micellar solutions that studied the interfacial dynamics and the
associated stress response
\cite{Becu.Manneville.ea04,Lopez-Gonzalez.Holmes.ea04,chaos2000,WFF98}.

In earlier work, Yuan \etal~\cite{Jupp.Yuan04} used a 2D finite
element algorithm to evolve a related JS model in planar shear.
Contrary to our results, they reported a stable interface. However,
they considered rather small values of $L_x$, perhaps in the stable
regime, and the finite element technique may have introduced
stabilising numerical diffusion.  Related recent work by
Onuki~\cite{Furukawa.Onuki05} on a two-fluid version of the DJS model
\cite{fielding03a} at imposed stress also showed a variety of
interesting time-dependent behaviour due to the interplay between
stress and concentration degrees of freedom; it would be interesting
to study the phenomena found here in that model. Indeed, an open
question is the extent to which the instability seen in this DJS model
is ubiquitous among other models of shear banding, such as those for
liquid crystals. Future work should also extend to larger systems and
smaller interfaces, likely to lead to more complex dynamical states.

SMF thanks Prof. Ajdari for his hospitality at the ESPCI in Paris
where this work was partly carried out, and UK EPSRC GR/S29560/01 for
financial support. We thank O. Harlen and H. Wilson for helpful
discussions.


\end{document}